# Cascade Generalization-based Classifiers for Software Defect Prediction

Aminat T. Bashir[1], Abdullateef O. Balogun[2], Matthew O. Adigun[3], Sunday A. Ajagbe[4], Luiz Fernando Capretz[5,6], Joseph B. Awotunde[7], Hammed A. Mojeed[7,8]

[1]Department of Computer Science, Al-Hikmah University, Ilorin 240281, Nigeria
[2]Department of Computer and Information Sciences, Universiti Teknologi PETRONAS, Seri Iskandar 32610, Malaysia
[3]Department of Computer Science, University of Zululand, Kwadlangezwa 3886, South Africa
[4]Department of Computer Engineering, First Technical University Ibadan, 200255, Nigeria
[5]Department of Electrical and Computer Engineering, Western University, London, ON N6A 5B9, Canada
[6]Division of Science, Yale-NUS College, Singapore 138533, Singapore
[7]Department of Computer Science, University of Ilorin, Ilorin, 1515 Ilorin, Nigeria
[8]Department of Technical Informatics and Telecommunications, Gdańsk University of Technology, Gabriela Narutowicza 11/12, 80-233 Gdańsk, Poland
*Corresponding author: abdullateef.ob@utp.edu.my

## Abstract

The critical importance of software testing to a successful software development process cannot be overstated. It is therefore imperative that software testing be conducted early and throughout the system's development phases to mitigate defective systems, especially given the limited availability of resources allotted to any given project. The process of software defect prediction (SDP) involves predicting which software system modules or components pose the highest risk of being defective. The projections and discernments derived from SDP can then assist the software development team in effectively allocating its finite resources toward potentially susceptible defective modules. Because of this, SDP models need to be improved and refined continuously. Hence, this research proposes the deployment of a cascade generalization (CG) function to enhance the predictive performances of machine learning (ML)-based SDP models. The CG function extends the initial sample space by introducing new samples into the neighbourhood of the distribution function generated by the base classification algorithm, subsequently mitigating its bias. Experiments were conducted to investigate the effectiveness of CG-based Naïve Bayes (NB), Decision Tree (DT), and k-Nearest Neighbor (kNN) models on NASA software defect datasets. Based on the experimental results, the CG-based models (CG-NB, CG-DT, CG-kNN) were superior in prediction performance when compared with the baseline NB, DT, and kNN models respectively. Accordingly, the average accuracy value of CG-NB, CG-DT, and CG-kNN models increased by +11.06%, +3.91%, and +5.14%, respectively, over baseline NB, DT, and kNN models. A similar performance was observed for the area under the curve (AUC) value with CG-NB, CG-DT, and CG-kNN recording an average AUC value of +7.98%, +26%, and +24.9% improvement over the baseline NB, DT, and kNN respectively. In addition, the suggested CG-based models outperformed the Bagging and Boosting ensemble variants of the NB, DT, and kNN models as well as existing computationally diverse SDP models.

## 1. Introduction

The persistent solicitation of system specifications from users often results in budget and schedule overruns in software development projects. Improving the completion rate of software has proven to be a major challenge encountered by the software development industry, leading to numerous efforts to address the issue [1]. Research indicates that there is a correlation between the budgeted amount for software development and the degree of complexity of the software. Additionally, it has been found that as project budgets increase, there is a temporary decrease in the success rate of the project [2]. Humphrey investigates the underlying causes of project failures and conducts a comprehensive analysis of the factors that should be considered for improving the efficacy of software projects on a large scale [3,4]. The study revealed that budgets below $750,000 yield a success rate of approximately 55%. However, as project size increases, success rates rapidly decline. According to existing studies, when the magnitude exceeds $10,000,000, the probability of success approaches nearly zero [3,4]. The aforementioned phenomenon is a striking actuality that poses a significant hindrance to the efficacy of enterprises engaged in software development.

The occurrence of defects in a software module can be attributed to invalid programming principles or erroneous code, resulting in inaccurate output and substandard software products [5]. The presence of defective software modules can lead to increased expenses in both development and maintenance and also leads to customer dissatisfaction, often resulting in consequences such as contract termination or revision [6]. The existence of a software defect, therefore, can result in substandard software and potentially contribute to the failure of a software project [7]. Software metrics are commonly employed to evaluate the efficacy and calibre of software products. Software engineers utilize software metrics to conduct risk assessments and accurately predict defects, thereby facilitating the identification and enhancement of software project quality [8].

The use of a Software Defect Prediction (SDP) process is thus thought highly beneficial during the software testing stages of the software development process [9,10]. The aforementioned source identifies modules that exhibit a higher likelihood of defects and therefore necessitate thorough testing [11]. By utilizing resources in this manner, it is possible to optimize their efficiency while adhering to the imposed limitations. While SDP is a valuable tool in the realm of software testing, accurately forecasting which modules will be defective is not always a straightforward task. Several challenges can hamper the seamless execution and utilization of SDP models [12].

Typically, an SDP model is constructed utilizing software metrics information that has been gathered from previously developed systems, releases, or analogous software endeavours [7]. Once the model has been validated, it is suitable for predicting the likelihood that program modules will be prone to faults during their development phase. The objective is to attain elevated levels of software dependability and excellence through efficient utilization of the existing resources. This study examines software metrics that pertain to the quality of software products and processes, as documented in sources [13-15] and have the potential to demonstrate and be represented as the attributes or features of software modules [8,16,17].

Therefore, the task of creating a software system that is free of defects is a challenging one, and it is common for there to be unidentified bugs or unanticipated shortcomings in developed systems, even when software process best practices are meticulously followed [18]. Several classification models based on diverse computational characteristics have been devised by researchers for SDP with comparative outcomes [19-21]. The reported findings from these studies suggested the application of ML classification algorithms for SDP tasks. However, the predictive performances of these SDP models, in most cases, remain unsatisfactory. This makes devising novel and scalable SDP models imperative. Hence, this study suggests the development of cascaded generalized (CG)-based classification models for SDP.

The working principle of CG is based on a multi-layered learning framework that allows base classifiers access to the main data features generated at the lower levels. That is, CG achieves coverage across the sample space by employing projections of the baseline classifier and appending these successive projections to each training instance as new models. As a result, both the base and meta-level models utilize the original sample space values, while the meta-level models also gain advantages from additional features represented by the projections obtained from the baseline model.

In summary, the following are the contributions of this research

I. Development of cascading generalization models for classification processes
II. Deployment of cascading generalization models using diverse ML algorithms for SDP tasks.
III. Empirical evaluation and validation of the predictive performances of the deployed cascading generalization models against baseline and current SDP models.

The remainder of this article is divided into the following sections. In Section 2, a thorough evaluation of relevant recent findings from existing studies is offered. The research technique and methods used in this study are explained in Section 3, along with a description of the suggested CG-based models and experimentation procedures. Section 4 presents the empirical results and the observed findings, while Section 5 offers a summary of the research and suggests directions for future research activities

## 2. Related Works

In this section, current SDP models or solutions from existing related studies will be examined. These models vary and are based on diverse computational characteristics. Specifically, SDP studies based on statistical, ML, and DL methods are analyzed.

It is important to emphasise that the Software Defect Prediction (SDP) approach is not the sole methodology available for identifying flaws or bugs within software systems. For example, the utilization of model checking [22] and static code analysis tools such as Coverity [23] can be employed to identify flaws in software systems. Model checking and static analysis are also widely used techniques for defect detection. These methods rely on fault localization to identify and pinpoint flaws within software systems. In this context, the discrepancies observed between the inputs of failed and successful software tests are utilised to detect and pinpoint issues within the source code. The conventional techniques commonly identify defects solely within the current codebase under examination, whereas SDP provides a proactive indication of potential defect-prone areas in a software system.

The initial approaches to SDP rely predominantly on the evaluation of software requirement specification (SRS) documents to identify potential deficiencies. An example of this can be seen in the work of Smidts, *et al.* [24], who devised a software reliability model that is grounded in the Software Requirements Specification (SRS) and failure methods. The input data for the proposed solution consists of the failure method'(s) data. In another study, Cortellessa, *et al.* [25] integrated the Unified Modelling Language (UML) of software architecture with a Bayesian framework for SDP. Regrettably, these approaches do not facilitate or enable the reuse of code, thereby implying that component failures are not interconnected. Gaffney and Davis [26,27] proposed a model for software reliability that operates on a phase-based framework. The model primarily relies on defect data that was identified after an inspection of various stages in software development. Nevertheless, this methodology is inherently limiting as it is typically customized to suit a specific institution. , Al-Jamimi [28] investigated the application of fuzzy logic in the context of SDP where the Takagi-Sugeno fuzzy inference engine was employed for SDP. Yadav and Similarly, Yadav [29] effectively implemented a fuzzy logic approach for the phase-wise software development process. The objective is to offer a comprehensive range of features for SDP. However, the question of decidability is a significant limitation of fuzzy logic [30].

To improve SDP solutions, some SDP models are based on statistical or discriminant analysis. For instance, [31] proposed a discrimination analysis based on the inherent characteristics of the defect datasets. They observed that the attributes of SDP datasets are mostly nonlinearly separable and imbalanced. Hence, they developed a novel kernel discrimination classifier (KDC) to build more efficient SDP models. Their results showed that KDC is effective and comparable to some existing SDP models. [32] proposed a novel method using a subclass discriminant analysis (ISDA) for SDP. ISDA was enhanced to tackle the latent class label disparity issue in within-project SDP and a semi-supervised transfer component analysis (SSTCA) method was developed for the cross-SDP. A robust SSTCA+ISDA prediction method was eventually designed to handle both within and cross-project SDP. Most of these statistical-based SDP models are easy to implement but they do not work well when the datasets are not inseparable and they are not as effective as machine learning (ML) or deep learning (DL)-based SDP models.

Regarding ML-based SDP models, [33] investigated the effectiveness of tree-based classifiers for SDP, based on the notion that tree-based classifiers are simple and effective in ML tasks. Ten tree-based classifiers with diverse computational characteristics were implemented on NASA datasets. Reports from their experimental results showed that RF is superior to other tree-based classifiers. In a similar study, [34] applied k-nearest neighbour (kNN), random forest (RF), and multilayer perceptron (MLP) on NASA. Their research aims to determine the effectiveness of the deployed ML classifiers for SDP. Findings from their results showed that kNN outperformed other experimental methods. However, it was observed that the scope of the research was limited based on the number of classifiers and datasets used. In addition, the parameter tuning of kNN is a major drawback since its performance is dependent on its parameter settings. To address the parameter setting problem of SDP models, [35] investigated the impact of parameter optimization on 26 SDP models of diverse computational characteristics on 18 defect datasets. Findings from the results indicated that the parameter optimization of ML classifiers can enhance their prediction performances and generate a more stable SDP model. However, while these ML classifiers were successfully implemented, they are often affected by misclassification and in some cases increased overhead costs for data preprocessing.

Another area gaining more attention and involvement is DL-based SDP models. [36] examined the application of DL methods for automatically predicting system vulnerability defects and proposed the application of DL models in automated software vulnerability defect prediction. In another study, [37] investigated the SDP performances of MLP and Convolutional Neural Networks (CNN). From their results, it was observed that although the DL models outperformed conventional ML models, modifying the parameters of the DL model directly influences their predictive performances. [38] proposed a novel method based on the hybridization of word embedding and DL methods for SDP. A semantic Long Short Term Memory (LSTM) network was developed that extracts features based on the word embedding model and thereafter deploys the LSTM model on the semantic data to conduct SDP. [39] developed gated hierarchical LSTM networks (GH-LSTMs). The results of these studies support the applicability of DL models for SDP. [40] developed a hybrid model based on CNN and Bi-LSTM. This model can predict the defective areas of source code by extracting AST tokens as vectors from source code, using CNN to extract the semantics of the AST tokens, and then deploying Bi-LSTM for the prediction process. In a similar study, [41] hybridized BiLSTM and BERT-based semantic features. The proposed method demonstrated SDP performances competitive with and comparable to existing state-of-the-art SDP models. In summary, these aforementioned DL models can generate effective SDP models with less dependency on data preprocessing. However, the DL processes are somewhat black-box and in most cases, it is difficult to understand the working principles of their internal mechanisms. Also, DL models are computationally expensive and heavily dependent on parameter tuning.

In summary, many ML and DL approaches and techniques for SDP have been proposed and developed. However, there is a continuous and urgent need to research and develop more accurate and sophisticated SDP models or methods. Consequently, this study suggests CG-based models for SDP.

## 3. Methodology

This section highlights the research and experimental procedure of this study. Specifically, the baseline ML classifiers, software datasets, proposed CG method, performance measures, and experimentation processes are analyzed.

### 3.1 Baseline Machine Learning Classifiers

Naïve Bayes (NB), DT, and kNN algorithms were selected as baseline classifiers in this research. The selection of these classifiers was based on their diverse computation characteristics for introducing heterogeneity into the experimental process. Additionally, the respective prominence and frequent utilization of these classifiers in existing SDP studies were some of the factors considered for their selection. For instance, NB utilizes Baye's theory in its prediction and is based on the assumption that instances are independent of themselves [42]. In the case of DT, it is an instance of a non-parametric classifier which generates a model or performs prediction using simple rules that are based on an initial value extracted from the dataset [43]. Lastly, kNN is an instance-based model that categorizes datasets based on a similarity method. As a lazy model, KNN does not consider underlying assumptions with its function approximated locally and full computation is postponed until classification [19]. Table 1 presents the parameter settings of the baseline classifiers as utilized in this research work.

**Table 1.** Parameter setting of the implemented classifiers.

| Classifiers | Parameter Setting |
|---|---|
| NB | UseKernelEstimator=True; UseSupervisedDiscretization=False |
| DT | ConfidenceFactor = 0.25; MinObj = 2 |
| KNN | K=1; distanceWeighting=False; NearestNeighbourSearchAlgorithm = LinearNNSearch; DistanceFunction=EuclideanDistance |

## 3.2 Cascade Generalization (CG) Classification Method

The fundamental concept of Cascade Generalization (CG) involves the sequential utilization of a series of classifiers. At each step, this process entails the expansion of the initial dataset through the incorporation of additional attributes. CG is an effective technique that can be used for both fundamental and advanced (meta-level) processes. In this study, CG is deployed as an advanced technique that uses a meta-level classifier to train and develop an efficient model explicitly based on the projections of the baseline classifier [44]. CG spans the density of the sample space by making use of the projections of the baseline classifier and attaching the successive projections of it to every training instance as a new model. Consequently, the base and meta-level models utilize the initial sample space values, with the meta-level models gaining benefits from extra features which are the projections from the baseline model [44].

For instance, Let D = {($x_a$, $y_b$), a,b = 1,… N} } be the training set for base-level classifiers, such that a = 1, 2, … , N is the number of instances, and $y_b$ ∊ Y= ($z_1$,$z_2$ ) is the class label, and the meta-dataset can be represented as newD= {(x, $p_1$, …, $p_{|B|}$, $y$)}, where B is the total number of baseline classifiers. For the baseline classifier, the conditional probability distributions $p_{1N}$ is replaced by the class probability. The final model is generated as a unified projection by amplifying the collected individual projections of the baseline models. The algorithm and the parameter values for the proposed CG-based method are depicted in Figure 3 and Table 2 respectively.

Algorithm 1. Cascade Generalization (CG) Method

*Training dataset D = {$\boldsymbol{x_a}$, $\boldsymbol{y_b}$}; a = 1, 2, … , N; $\boldsymbol{y_b}$ ∊ Y= ($\boldsymbol{z_1}, \boldsymbol{z_2}$ ) $\boldsymbol{z_1}$ is the class label;*
*Iterations I = 100*
*Baseline Model B = (NB, DT, kNN)*
*D′ = ExtendDataset*
Begin
ExtendDataset (i, D)
newD = Ø
for each $\alpha$ = (x,y) ∊ D
    for k= 1 to |Y|
        $\boldsymbol{p_k = probability\ that\ \alpha}$ ∊ $\boldsymbol{y_k}$ according to *i*
    $\alpha'$ = (x, $\boldsymbol{p_1}$, …, $\boldsymbol{p_{|y|}}$, $\boldsymbol{y}$)
    newD = newD ∪ { $\alpha'$}
return newD
CG ({$\boldsymbol{B_1}$, $\boldsymbol{B_2}$, $\boldsymbol{B_3}$ , D)
for j = 1 to N
$\boldsymbol{i_j}$ = model induced by $\boldsymbol{B_j}$ from D

$\boldsymbol{D_j}$= D′ ($\boldsymbol{i_j}, \boldsymbol{D}$)
For each new subset of instance s
S′ = D′ ($\boldsymbol{I_j}$, {s} )
label (s) = $\boldsymbol{i_2}$ ($\boldsymbol{s'}$)
End

**Figure 3:** Proposed CG-based method

**Table 2.** Parameter setting of the Cascade Generalization Method

| Classifiers | Parameter Setting |
|---|---|
| CG | Classifier= {DT, NB, kNN}; ContatenatePredictions=True; KeepOriginal=True; meta=RF; numIterations=10 |

## 3.3 Software Defect Datasets

In this research, the software metrics from the NASA repositories were used for the training and testing of the investigated models. In particular, we used the [45] version of the NASA corpus for our experiments. The software metrics in the NASA repositories are based on static code assessments [46-48]. Further descriptions of the software metric datasets are presented in Table 3.

Table 3. Software Metric Datasets

| Datasets | Instances | Features | Defective Instances | Non-Defective Instances |
|---|---|---|---|---|
| CM1 | 327 | 38 | 42 | 285 |
| KC1 | 1126 | 22 | 294 | 868 |
| KC3 | 194 | 40 | 36 | 158 |
| MC2 | 124 | 40 | 44 | 80 |
| MW1 | 250 | 38 | 25 | 225 |
| PC1 | 679 | 38 | 55 | 624 |
| PC3 | 1053 | 38 | 130 | 923 |
| PC4 | 1270 | 38 | 176 | 1094 |
| PC5 | 1694 | 39 | 458 | 1236 |

## 3.4 Experimental Framework

This section presents and explains the experimental procedure conducted in this research. Specifically, Figure 4 depicts a schematic illustration of the experimental procedures carried out in the present research.

The framework depicted in Figure 4 is developed to evaluate and verify the effectiveness of the suggested methods through empirical evaluation. The study employs an experimental structure to analyze software metric datasets from the NASA repository. The cross-validation (CV) method is used for the development of training and testing of models. CV was utilized in this research based on its ability to produce predictive models with reduced bias and variance, as evidenced by prior research [49-54]. Additionally, the CV methodology involves the incremental utilization of each instance within a dataset for both training and testing purposes. The CV technique has been elucidated in the literature, specifically in references [16,46,55-58]. The SDP performances of the suggested CG-based models are initially compared with the baseline classifiers to ascertain their effectiveness. Furthermore, the SDP performances of the CG-based models were compared with the Bagging and Boosting ensemble variants of the baseline classifiers. All models were implemented using the machine learning libraries from WEKA[43].

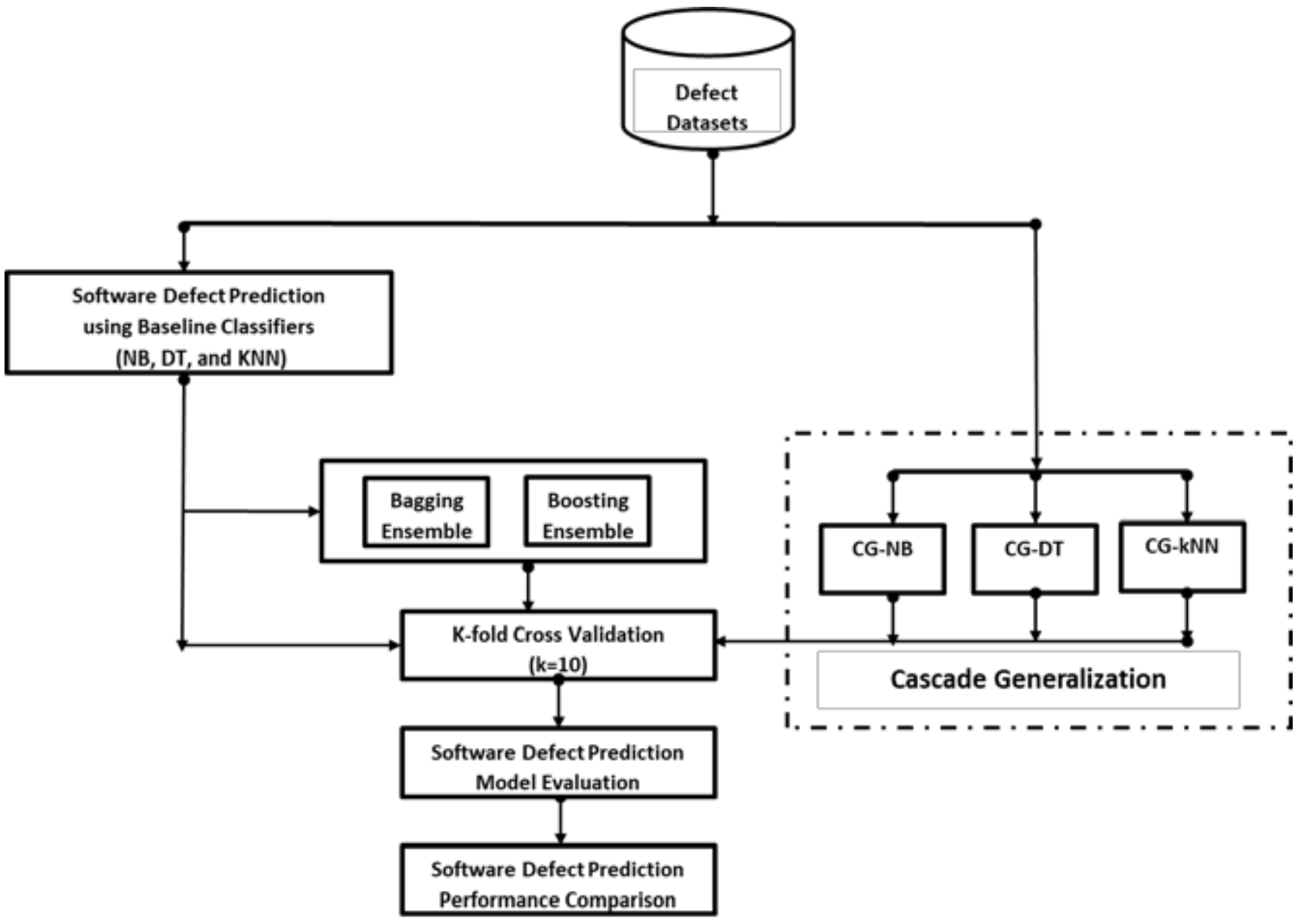

Figure 2. Experimental Framework

### 3.5 Performance Assessment Measures

This research employed well-established evaluation metrics, including accuracy, f-measure, the area under the curve (AUC), and Mathew's Correlation Coefficient (MCC), to assess and contrast the predictive capabilities of various models. The selection of these performance indicators was based on their frequent utilization in prior studies for the assessment of rule-based and ML-based software risk prediction models [16,59,60]. Furthermore, these metrics are reported to be dependable collectively as they consider all areas of the confusion matrix produced for each model developed [61,62].

## 4. Experimental Results and Discussion

For clarity, the analysis of the experimental results presented in this section is divided into three parts as outlined in the succeeding subsections.

**4.1** Scenario 1: Experimental Results of CG-based Models and the Baseline Classifiers

This sub-section presents the experimental results of the CG-based models we developed and the baseline classifiers, it is important to indicate the effectiveness of the suggested CG-based models in comparison to the prominent baseline classifiers. Consequently, the SDP capabilities of the CG-based models and baseline classifiers on the studied defect datasets, based on accuracy and AUC values, are presented and discussed in Tables 4 and 5 respectively.

Table 4. Accuracy values of CG-based Models and Baseline Classifiers

| Accuracy | NB | CG-NB | DT | CG-DT | KNN | CG-KNN |
|---|---|---|---|---|---|---|
| CM1 | 81.34 | 84.71 | 81.04 | 85.32 | 77.98 | 85.32 |
| KC1 | 73.58 | 77.45 | 74.18 | 76.85 | 73.23 | 77.62 |
| KC3 | 78.86 | 78.35 | 79.38 | 79.90 | 72.16 | 80.41 |
| MC2 | 70.16 | 70.97 | 60.48 | 71.77 | 70.16 | 70.97 |
| MW1 | 81.60 | 90.00 | 90.40 | 90.00 | 83.60 | 90.40 |
| PC1 | 89.10 | 92.19 | 91.45 | 92.05 | 90.72 | 91.90 |
| PC3 | 39.41 | 87.84 | 84.91 | 87.94 | 84.90 | 86.99 |
| PC4 | 85.35 | 90.16 | 86.92 | 88.98 | 85.66 | 89.06 |
| PC5 | 74.97 | 77.33 | 73.90 | 78.16 | 74.14 | 76.45 |
| Average | 74.93 | 83.22 | 80.30 | 83.44 | 79.17 | 83.24 |

As shown in Table 4, the CG-based models had higher accuracy values on the experimented defect datasets, as high as 90% and as low as 70% across the datasets. This finding indicates the high SDP performances of the CG-based models. In most cases, the CG-based models recorded superior prediction accuracy values when compared to the baseline classifiers. For instance, on the PC3 dataset, CG-NB (87.84%) had more than a +100% increment in its accuracy value over the NB model (39.41%). Similar results were observed in the case of CG-DT and CG-kNN on the same PC3 dataset. CG-DT and CG-kNN had accuracy values of 87.94% and 86.99% respectively which are +3.44% and +2.46% higher than the accuracy values of DT (84.90%) and kNN (84.91%).

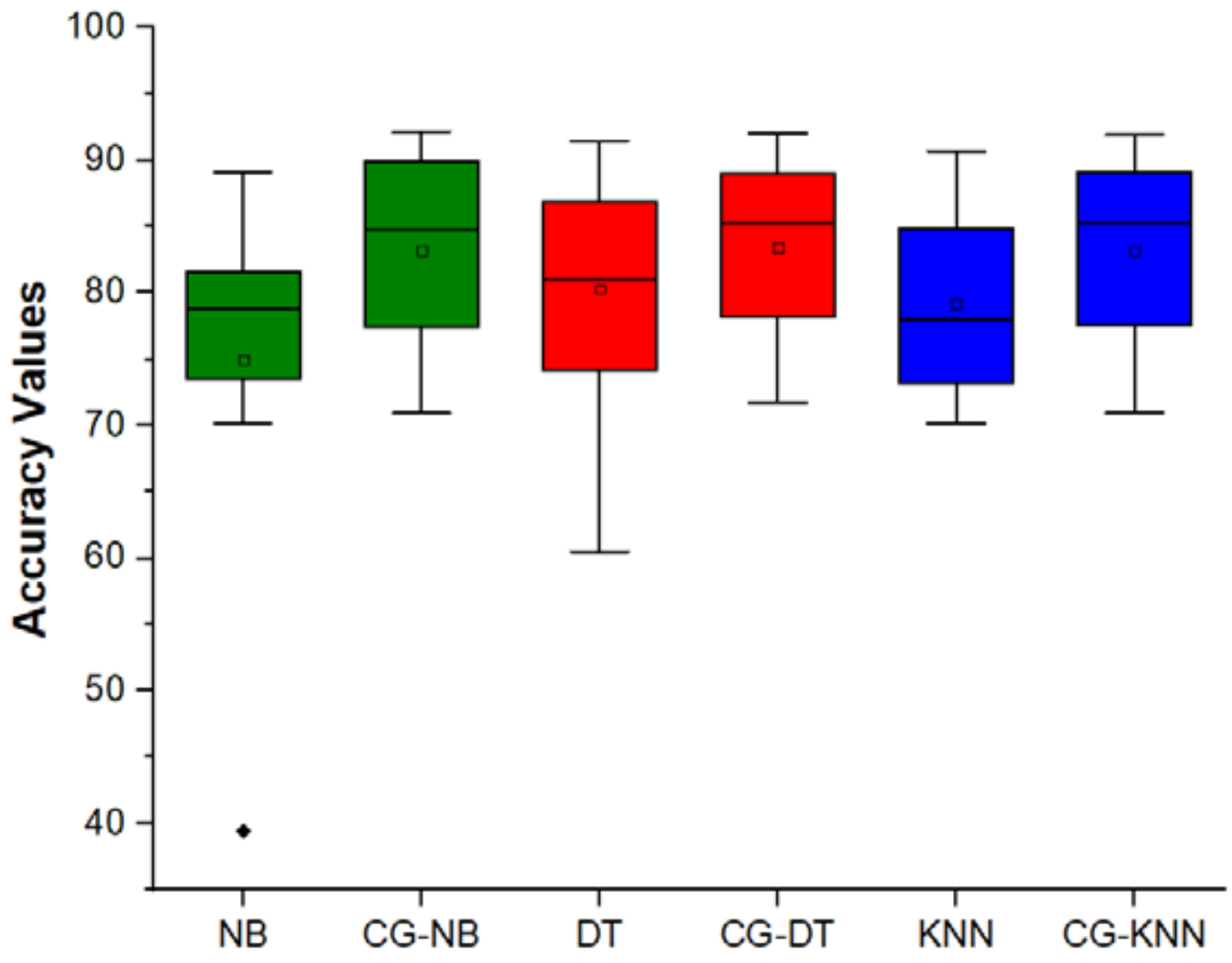

Figure 5. CG-based model and baseline classifier accuracy value box plot representation

Also, on Dataset CM1, CG-NB (84.71%), CG-DT (85.32%), and CG-kNN (85.32%) had a +4.14%, +5.28%, and +9.41% increment in their respective accuracy values over NB, DT, and kNN models.. In addition, CG-based models (CG-NB:83.22%, CG-DT: 83.44%, CG-kNN: 83.24%) had better accuracy values compared to the baseline classifier models (NB:74.93%, DT: 80.30%, kNN: 79.17%). CG-based model and baseline classifier accuracy value box plot representation are presented in Figure 5.

Table 5. AUC values of CG-based Models and Baseline Classifiers

| AUC | NB | CG-NB | DT | CG-DT | KNN | CG-KNN |
|---|---|---|---|---|---|---|
| CM1 | 0.645 | 0.704 | 0.570 | 0.717 | 0.521 | 0.689 |
| KC1 | 0.681 | 0.732 | 0.604 | 0.723 | 0.633 | 0.733 |
| KC3 | 0.662 | 0.723 | 0.653 | 0.731 | 0.539 | 0.703 |
| MC2 | 0.707 | 0.702 | 0.589 | 0.697 | 0.653 | 0.718 |
| MW1 | 0.778 | 0.731 | 0.503 | 0.723 | 0.607 | 0.724 |
| PC1 | 0.791 | 0.885 | 0.598 | 0.861 | 0.679 | 0.864 |
| PC3 | 0.749 | 0.846 | 0.591 | 0.834 | 0.603 | 0.830 |
| PC4 | 0.814 | 0.937 | 0.789 | 0.925 | 0.700 | 0.935 |
| PC5 | 0.719 | 0.803 | 0.673 | 0.806 | 0.667 | 0.800 |
| Average | 0.727 | 0.785 | 0.619 | 0.780 | 0.622 | 0.777 |

As observed in Table 5, the CG-based models had high AUC values on the defect datasets. The CG-based models had AUC values as high as 0.937 and as low as approximately 0.700 across the datasets. However, in most cases, in comparison to the baseline classifiers, the CG-based models still achieved superior AUC values to the baseline classifiers. This finding indicates that the CG-based model performed well and has a high degree of separability between the defective and non-defective instances in the defect datasets.

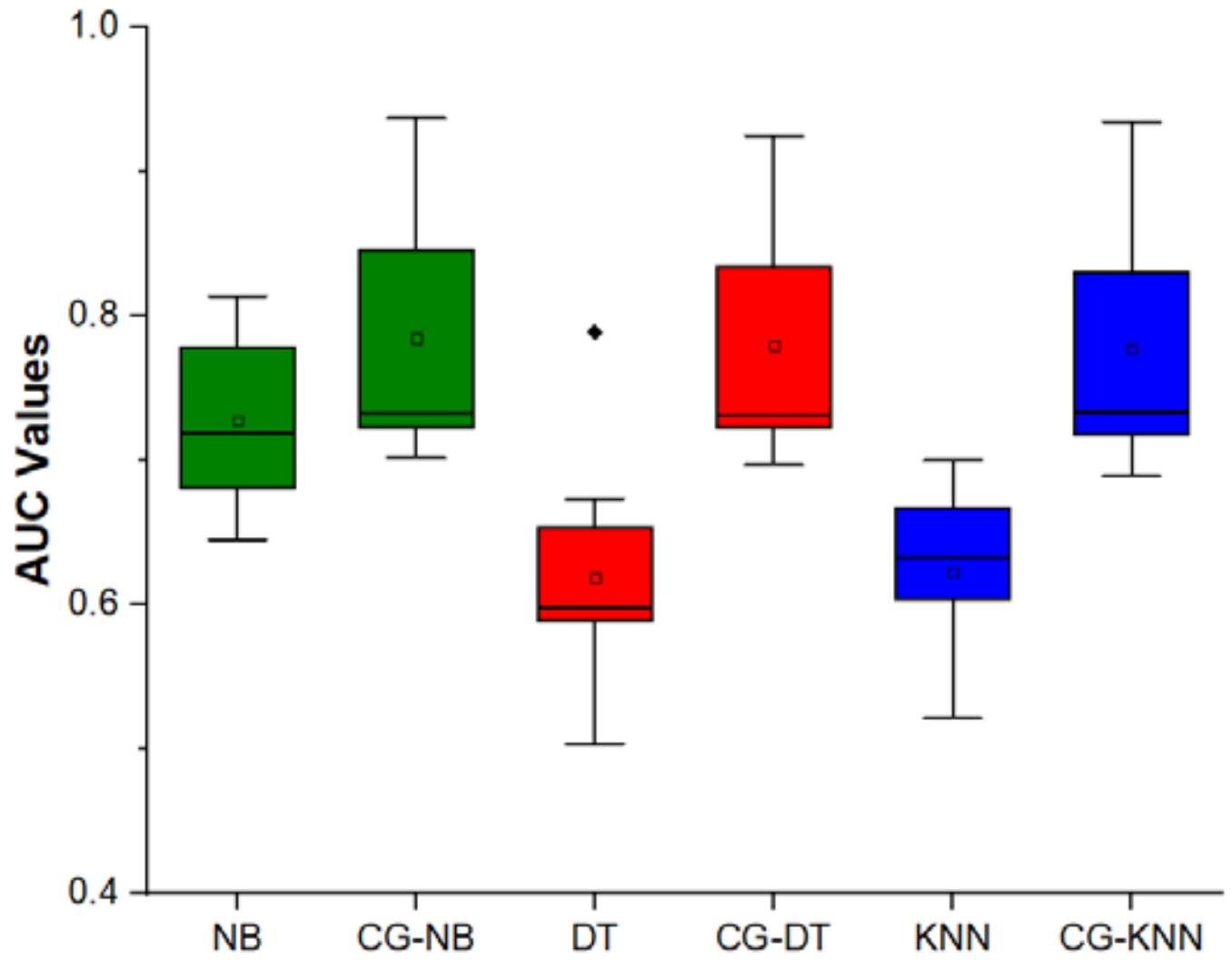

Figure 6. CG-based model and baseline classifier AUC value box plot representation

As in the case of dataset PC4, CG-NB (0.937) outperformed NB (0.814) with a +15.11% increment in its AUC value. Similarly, CG-DT and CG-kNN outperformed DT and kNN models respectively on the same PC4 dataset, with AUC values of 0.925 and 0.935 respectively, which are +17.24% and +33.57% increments over the AUC values of DT (0.789) and kNN (0.700) respectively. Furthermore, on Dataset PC3, CG-NB (0.846), CG-DT (0.834), and CG-kNN (0.830) had a +12.95%, +41.12%, and +32.67% increment in their respective AUC values over NB, DT, and kNN models on the same PC3 dataset. The high variation in the AUC values of the CG-based models over their baseline counterparts may be attributed to the skewness in the class labels of the experimental datasets as depicted in Table 3. In other words, CG-based models can accommodate the latent imbalance problem better than the single baseline classifiers. On average, CG-based models (CG-NB:0.785, CG-DT: 0.780, CG-kNN: 0.777) had superior AUC values compared to the baseline classifier models (NB:0.727, DT: 0.619, kNN: 0.622). Notably, the SDP performance of the baseline classifiers was commensurate in some cases, whereas the SDP performance of the CG-based models is superior on average. The CG-based model and baseline classifier AUC value box plot representation are presented in Figure 6.

### **4.2** Scenario 2: Experimental Results of CG-based Models and the Ensemble Methods

Table 3 presents the predictive performance comparisons of FURIA and its nested subset selection dichotomy variants (FURIA-BFS, FURIA-BSS, and FURIA-RPS) on the studied dataset. As indicated, the suggested FURIA variants outperformed the conventional FURIA model. Specifically, FURIA-FCS had the highest accuracy (98.62%), f-measure (0.987), and MCC (0.983) values with +0.64%, +0.71%, and +0.82% increments over the predictive performance of FURIA on the same dataset. The duo of FURIA-BSS and FURIA-RPS had similar predictive performances of 98.33% accuracy value, 0.983 f-measure value, and 0.979 MCC values. The differences in the predictive performances of the suggested fuzzy induction models may seem insignificant, but the cost of misprediction or misclassification is dire.

For a more generalizable empirical SDP performance comparison, the predictive performances of the CG-based models are further compared with the ensemble variants of the baseline models. Specifically, the SDP performances of Boosting (Bo) and Bagging (Bg) variants of NB, DT, and kNN models are compared with the CG-based models. The succeeding subsection presents a detailed experimental analysis of this comparison. The goal of this analysis is to further ascertain and confirm the effectiveness of the proposed CG-based models with ensemble methods. Existing studies have shown that ensemble methods are superior in predictive performances to individual baseline models.

Tables 6 and 7 present the SDP performances of the proposed models and the ensemble variants of the baseline classifiers based on accuracy and AUC values respectively.

**Table 6.** Accuracy values of CG-based Models and Ensemble variants of Baseline Classifiers

| Accuracy | Bg-NB | Bo-NB | CG-NB | Bg-DT | Bo-DT | CG-DT | Bg-DT | Bo-DT | CG-kNN |
|---|---|---|---|---|---|---|---|---|---|
| CM1 | 81.65 | 81.34 | 84.71 | 85.01 | 83.79 | 85.32 | 77.98 | 77.98 | 85.32 |
| KC1 | 73.92 | 74.10 | 77.45 | 77.54 | 73.32 | 76.85 | 74.18 | 73.24 | 77.62 |
| KC3 | 77.32 | 79.90 | 78.35 | 84.54 | 79.38 | 79.90 | 74.74 | 72.16 | 80.41 |
| MC2 | 70.16 | 69.35 | 70.97 | 66.13 | 66.94 | 71.77 | 71.77 | 70.16 | 70.97 |
| MW1 | 82.00 | 85.60 | 90.00 | 89.20 | 88.80 | 90.00 | 83.20 | 83.60 | 90.40 |
| PC1 | 87.92 | 90.57 | 92.19 | 91.46 | 91.31 | 92.05 | 90.72 | 90.72 | 91.90 |
| PC3 | 48.62 | 39.41 | 87.84 | 85.94 | 85.66 | 87.94 | 84.33 | 84.90 | 86.99 |
| PC4 | 84.65 | 85.83 | 90.16 | 89.13 | 88.19 | 88.98 | 85.35 | 85.67 | 89.06 |
| PC5 | 74.97 | 74.91 | 77.33 | 76.80 | 74.73 | 78.16 | 73.97 | 74.14 | 76.45 |
| Average | 75.69 | 75.67 | 83.22 | 82.86 | 81.35 | 83.44 | 79.58 | 79.17 | 83.24 |

Table 6 shows that the CG-based models outperform the ensemble variations of the baseline models in terms of accuracy values. This experimental finding can be observed across the datasets we compared, except for datasets KC1, KC3, and PC4, where Bo-DT recorded the best accuracy values. For instance, on the CMI dataset, CG-NB and CG-kNN significantly outperformed the respective ensemble variants of Bg-NB, Bo-NB, Bg-kNN, and Bo-kNN respectively. Both CG-NB and CG-kNN had an increment of +3.74%, +4.14%, +9.41%, and +9.41% in accuracy values respectively. A comparable scenario may be observed in the context of the MW1 dataset where CG-NB and CG-kNN significantly outperformed the respective ensemble variants of Bg-NB, Bo-NB, Bg-kNN, and Bo-kNN respectively. On average, the suggested CG-based models are superior to the experimented ensemble variants of the base models. CG-NB (83.22%) had an average increment of +9.95% and +9.98% in accuracy values over Bo-NB (75.67%) and Bg-NB (75.69%). Similarly, CG-kNN recorded an average increment of +5.14% and +4.6% in accuracy values over Bo-kNN (79.17%) and Bg-kNN (79.58%). These findings further demonstrate the superiority of the CG-based models in SDP to the prominent ensemble variants of NB, DT, and kNN models. Figure 7 presents the box-plot representation of the accuracy values of the suggested CG-based models and the ensemble variants of the baseline classifiers.

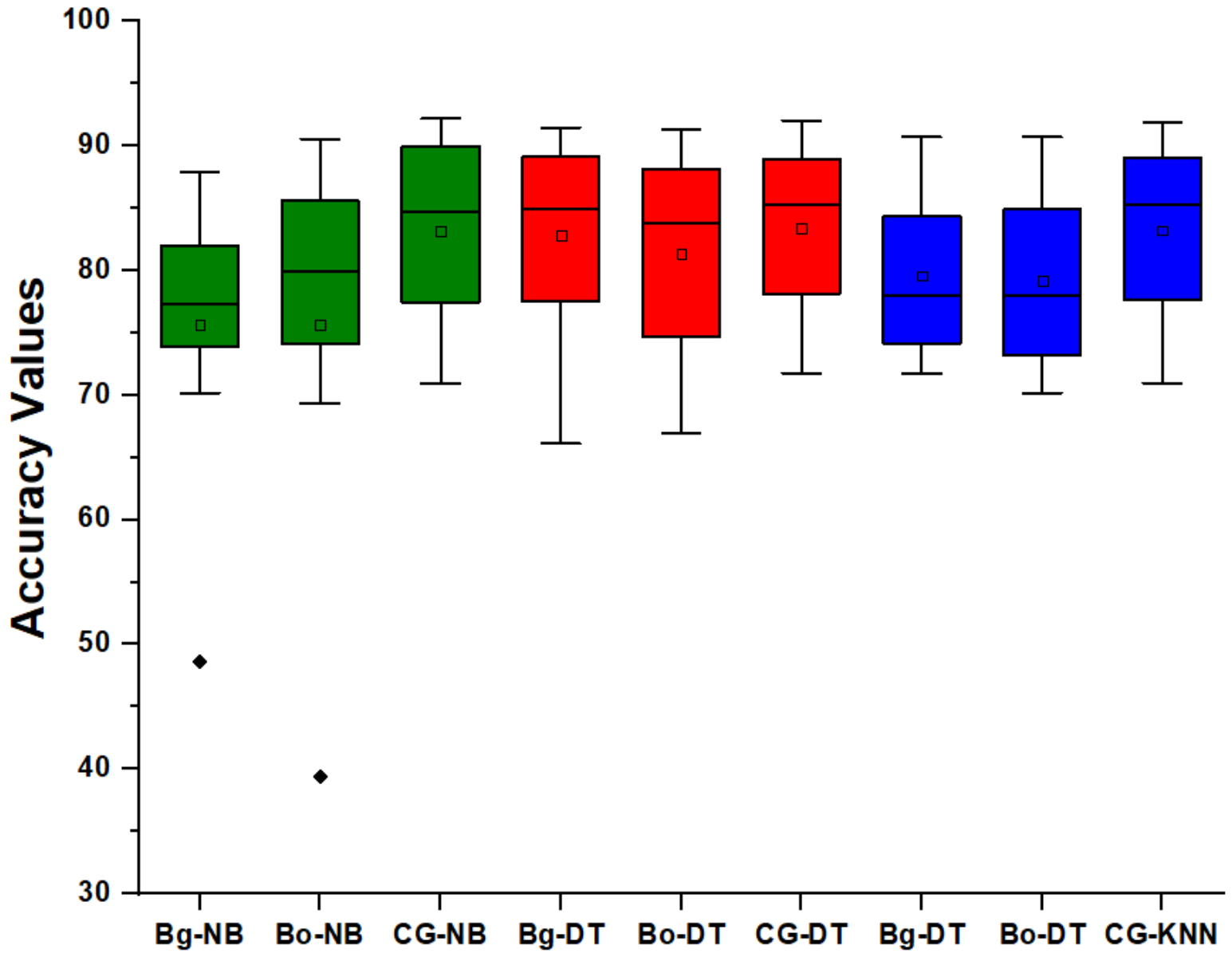

**Figure 7.** CG-based model and Ensemble variants of baseline classifiers accuracy value box plot representation

Table 7. AUC values of CG-based Models and Ensemble variants of Baseline Classifiers

| AUC | Bg-NB | Bo-NB | CG-NB | Bg-DT | Bo-DT | CG-DT | Bg-DT | Bo-DT | CG-kNN |
|---|---|---|---|---|---|---|---|---|---|
| CM1 | 0.664 | 0.623 | 0.704 | 0.733 | 0.736 | 0.717 | 0.593 | 0.521 | 0.689 |
| KC1 | 0.678 | 0.662 | 0.732 | 0.719 | 0.691 | 0.723 | 0.672 | 0.633 | 0.733 |
| KC3 | 0.674 | 0.650 | 0.723 | 0.729 | 0.712 | 0.731 | 0.605 | 0.539 | 0.703 |
| MC2 | 0.704 | 0.676 | 0.702 | 0.736 | 0.701 | 0.697 | 0.731 | 0.653 | 0.718 |
| MW1 | 0.772 | 0.774 | 0.731 | 0.749 | 0.715 | 0.723 | 0.678 | 0.607 | 0.724 |
| PC1 | 0.785 | 0.817 | 0.885 | 0.834 | 0.780 | 0.861 | 0.723 | 0.679 | 0.864 |
| PC3 | 0.738 | 0.741 | 0.846 | 0.788 | 0.788 | 0.834 | 0.700 | 0.603 | 0.830 |
| PC4 | 0.808 | 0.821 | 0.937 | 0.914 | 0.894 | 0.925 | 0.780 | 0.700 | 0.935 |
| PC5 | 0.715 | 0.711 | 0.803 | 0.789 | 0.772 | 0.806 | 0.723 | 0.667 | 0.800 |
| Average | 0.726 | 0.719 | 0.785 | 0.777 | 0.754 | 0.780 | 0.689 | 0.622 | 0.777 |

In addition, Table 7 highlights the AUC values of the CG-based models and the ensemble variants of the baseline classifiers. As shown in Table 6, the CG-based models recorded, on average, superior AUC values on the defect datasets we compared over the investigated ensemble variants. Specifically, the CG-NB had +8.13% and +9.18% average increments in its AUC values over Bg-NB and Bo-NB. Also, CG-kNN achieved an average increment of +12.77% and +24.92% over Bg-kNN and Bo-kNN models on the experimented datasets. This finding further illustrates that the CG-based models are effective and capable of differentiating the class labels.

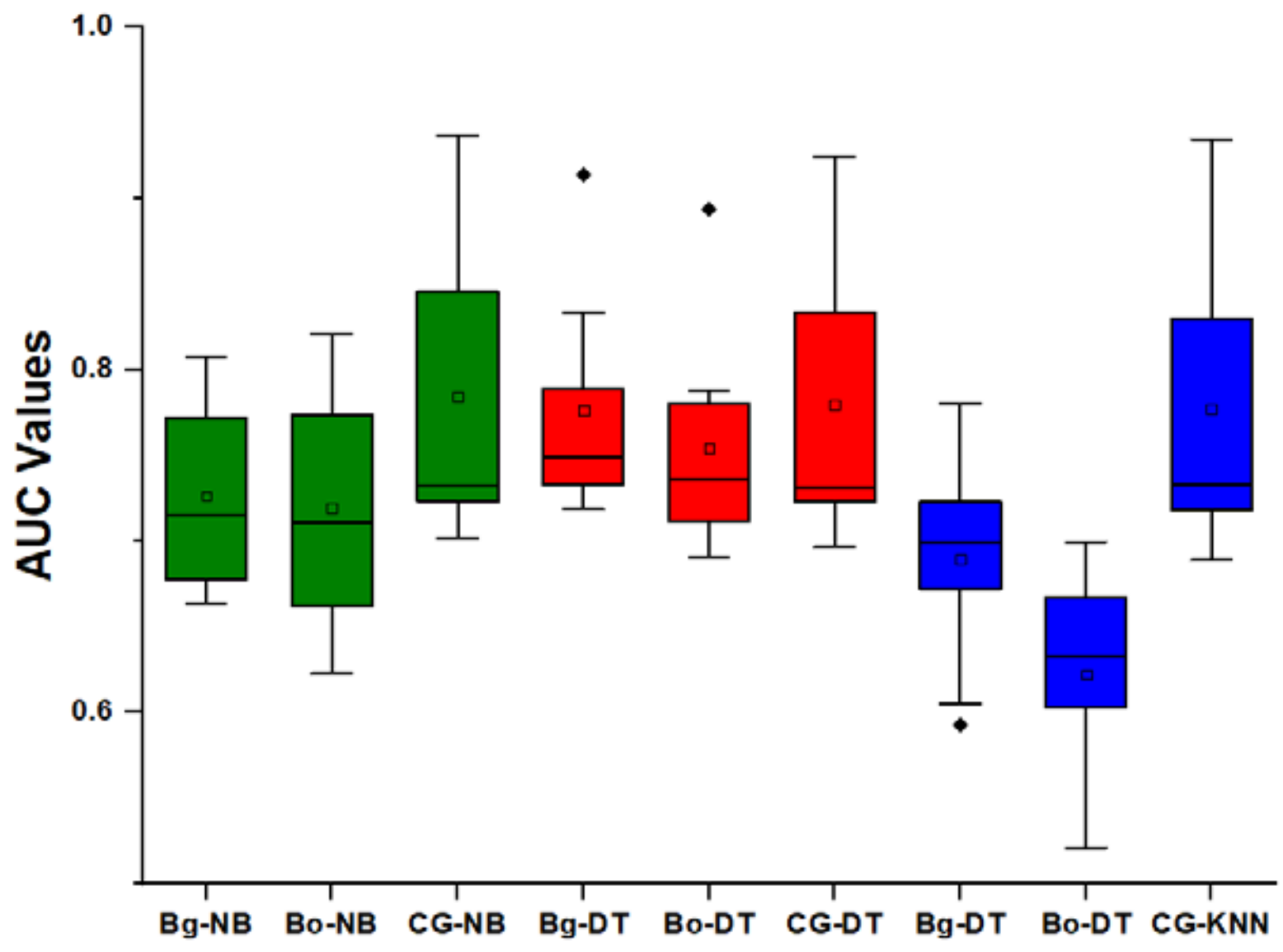

**Figure 8.** CG-based model and Ensemble variants of baseline classifiers AUC value box plot representation

Furthermore, similar to the findings from Table 6, based on the AUC values, the CG-based models outperformed the ensemble variations we investigated in most circumstances. This occurrence can be observed across the experimental datasets, except in the cases of datasets CM1, MC2, and MW1 where Bo-DT recorded the superior AUC values. For instance, on the CMI dataset, CG-NB and CG-kNN significantly outperformed the respective ensemble variants of Bg-NB, Bo-NB, Bg-kNN, and Bo-kNN respectively. Both CG-NB and CG-kNN had an increment of +3.74%, +4.14%, +9.41%, and +9.41% in AUC values respectively. These findings further showed the superiority of the CG-based models in SDP to the prominent ensemble variants of NB, DT, and kNN models. It is worth noting that the ensemble variants, particularly the Bg-DT, had competitive SDP performances in some cases, however, on average, the CG-based models are superior in SDP performance.. Figure 8 presents the box-plot representation of the AUC values of the suggested CG-based models and the ensemble variants of the baseline classifiers.

### **4.3** Scenario 3: Performance Comparison of CG-based Models and Existing SDP Models

In this subsection, the SDP performances of the proposed CG-based models are compared with the existing SDP models on the same metrics and defect datasets.

Table 8. Accuracy Value Comparison of CG-based Models with Existing SDP Methods

| SDP Models | CM1 | KC1 | KC3 | MC2 | MW1 | PC1 | PC3 | PC4 | PC5 |
|---|---|---|---|---|---|---|---|---|---|
| *CG-NB | 84.71 | 77.45 | 78.35 | 70.97 | 90.00 | 92.19 | 87.84 | 90.16 | 77.33 |
| *CG-DT | 85.32 | 76.85 | 79.90 | 71.77 | 90.00 | 92.05 | 87.94 | 88.98 | 78.16 |
| *CG-kNN | 85.32 | 77.62 | 80.41 | 70.97 | 90.40 | 91.90 | 86.99 | 89.09 | 76.45 |
| BaggedLR[63] | 74.00 | - | 76.00 | 65.00 | - | 81.00 | 75.00 | 83.00 | 68.00 |
| AdaboostSVM[63] | 75.00 | - | 77.00 | 65.00 | - | 79.00 | 74.00 | 81.00 | 68.00 |
| kStar[64] | 77.55 | 72.20 | 75.86 | 59.46 | 82.67 | 86.27 | 82.59 | 81.89 | 69.88 |
| CS-Forest[33] | 82.53 | - | 81.44 | - | 88.33 | 91.16 | 84.77 | 88.88 | - |
| Rotation Tree [33] | 83.33 | - | 70.61 | - | 86.60 | 91.07 | 85.54 | 86.69 | - |
| Dagging_NB[65] | 70.80 | 66.90 | 70.20 | 68.40 | 75.30 | 78.60 | 78.50 | 81.60 | 69.90 |
| Dagging_DT[65] | 59.60 | 68.10 | 64.50 | 70.80 | 77.10 | 76.70 | 78.20 | 89.70 | 76.50 |
| Dagging_kNN[65] | 61.10 | 67.90 | 62.30 | 70.60 | 72.20 | 78.50 | 75.30 | 84.40 | 76.40 |

* indicates the method proposed in this research

Table 8 compares the proposed CG-based models (CG-NB, CG-DT, and CG-kNN) to existing approaches in terms of prediction accuracy values. Specifically, the experimental results from existing studies such as [33], [63], [64], and [65] are compared with the suggested models on accuracy values. These existing models are based on diverse computational techniques and are reported to have had good predictive performances. For instance, the models suggested by [63] are based on homogeneous ensemble methods. Specifically, BaggedLR and AdaboostSVM methods were used for SDP. Also [66] developed a heterogeneous stacking ensemble method for SDP. As demonstrated in Table 8, the suggested CG-based models outperformed these current approaches in most cases (in most of the datasets). Likewise, existing methods based on enhanced instance learning, such as kStar [64], and enhanced forest (tree)-based learning, such as CS-Forest and Rotation Forest, have also had comparable prediction accuracy performances [33]. The suggested methods were also superior to the more recent approach based on dagging meta-learner as proposed by [65]. This finding verifies the efficacy and usability of CG-based SDP approaches.

Table 9. AUC value Comparison of CG-based Models with existing SDP methods

| SDP Models | CM1 | KC1 | KC3 | MC2 | MW1 | PC1 | PC3 | PC4 | PC5 |
|---|---|---|---|---|---|---|---|---|---|
| *CG-NB | 0.704 | 0.732 | 0.723 | 0.702 | 0.731 | 0.885 | 0.846 | 0.937 | 0.803 |
| *CG-DT | 0.717 | 0.723 | 0.731 | 0.697 | 0.723 | 0.861 | 0.834 | 0.925 | 0.806 |
| *CG-kNN | 0.689 | 0.733 | 0.703 | 0.718 | 0.724 | 0.864 | 0.830 | 0.935 | 0.800 |
| BaggedLR[63] | 0.650 | - | 0.660 | 0.610 | - | 0.770 | 0.740 | 0.870 | 0.680 |
| AdaboostSVM[63] | 0.680 | - | 0.660 | 0.590 | - | 0.760 | 0.730 | 0.820 | 0.680 |
| Stacking (NB, MLP, J48)[66] | - | - | - | - | - | 0.749 | - | - | - |
| Stacking (NB, MLP, J48)+SMOTE[66] | - | - | - | - | - | 0.871 | - | - | - |
| J48[67] | 0.594 | 0.689 | - | - | - | 0.668 | - | - | - |
| kStar[64] | 0.538 | 0.651 | 0.528 | 0.510 | 0.543 | 0.673 | 0.749 | 0.734 | 0.629 |
| Dagging_NB[65] | 0.708 | 0.669 | 0.702 | 0.684 | 0.753 | 0.786 | 0.785 | 0.816 | 0.699 |
| Dagging_DT[65] | 0.596 | 0.681 | 0.645 | 0.708 | 0.771 | 0.767 | 0.782 | 0.897 | 0.765 |
| Dagging_kNN[65] | 0.611 | 0.679 | 0.623 | 0.706 | 0.722 | 0.785 | 0.753 | 0.844 | 0.764 |

* indicates the method proposed in this research

For further comparison, Table 9 presents the comparative predictive performances of the proposed CG-based models (CG-NB, CG-DT, and CG-kNN) to existing approaches in terms of AUC values. Similar to the observed findings in Table 8, the proposed CG-based models still had superior AUC values to methods based on homogeneous ensemble methods [63], heterogeneous ensemble methods [66], tree methods [67], instance learning methods [64], and dagging meta-learner-based models [65].

## 5. Threat to Validity

Constraints observed in the current research, those documented in prior research, include the following: Two distinct categories of threats pose an obstacle to the validity of any given study: The potential threat to the generalizability of findings beyond the study's sample, also known as external validity, and the potential for confounding variables to affect the accuracy of causal inferences, also known as internal validity. The relative importance of external validity threats is considered to be higher than that of internal validity threats. The potential for making inferences between independent and dependent variables presents a challenge to internal validity, as noted in [68,69]. There exists a possibility that the data may not be cumulative, resulting in a gap between the numerical values that necessitate appropriate handling. The potential for external validity threats is associated with the extent to which the projected models can be applied to other contexts or populations. The results in this study were obtained through the utilization of the open-source software WEKA tool. Therefore, it is possible that the outcomes may not be generalizable to other systems. The efficacy of dependability prediction models is contingent upon the operating environment. Nevertheless, the dataset is not of considerable magnitude. The mitigation of these risks could potentially be achieved through the execution of additional replicated experiments across multiple platforms. Despite the various limitations and constraints, the outcome of our investigation offers guidance for future studies on the impact of previous dataset failures on software reliability forecasting through the utilization of machine learning techniques.

## 6. Conclusion and Future Works

The use of ML methods in SDP is now recognized as a promising new field of study. While it may be difficult to find defects in software at the beginning of software development, doing so may help ensure that high-quality products are delivered. Several SDP models have been proposed and developed by software experts and researchers. However, there is a continuous need for effective and sophisticated methods due to the rapid increase tn the volume of the modern software code base and interdependencies between the software modules and components. Our research proposed and deployed a CG-based method for SDP on NASA software defect datasets. The proposed CG-based models outperformed the investigated baseline classifiers of NB, DT, and kNN and their respective Bagging and Boosting ensemble variants on the NASA datasets studied. Specifically, the CG-NB model exhibited an average accuracy of 83.22%, surpassing the baseline Naive Bayes (NB) model's performance of 74.93% by a margin of 11.06%. Similar occurrences were observed with CG-DT and CG-kNN, with average accuracy values of 83.44% and 83.24%, which are +3.91% and +5.14% increments over DT and kNN respectively. Likewise, it was observed that CG-NB exhibited an average AUC value of 0.785, indicating a superiority of +7.98% over the baseline NB average AUC value score of 0.727. Furthermore, CG-DT and CG-kNN recorded an average AUC value of 0.780 and 0.777, indicating a +26% and +24.9% improvement over the baseline DT and kNN with an AUC value of 0.619 and 0.622 respectively. The predictive performances of the suggested models were superior to most of the current computationally diverse SDP models. Hence, it is recommended that we consider the CG-based method for SDP and ML tasks as it outperformed and can be comparable in predictive performance with existing ML models.

To further advance this research, more experimentation on other forms of defect datasets, such as the PROMISE repository, will be conducted. Additionally, the issue of data quality problems, particularly the high dimensionality and the class problem, will be explored and an appropriate ML-based model that can accommodate these data quality problems will be developed.